\documentclass[final,twocolumn]{elsarticle}

\usepackage{graphics}

\usepackage{amssymb}

\usepackage{array}

\journal{archive}

\begin{document}
\begin{frontmatter}

\title{Gas damping force noise on a macroscopic test body in an infinite 
gas reservoir}


\author{A.~Cavalleri,$^a$ G.~Ciani,$^b$ R.~Dolesi,$^b$  
M.~Hueller,$^b$ D. Nicolodi,$^b$ D. Tombolato,$^b$ 
S.~Vitale,$^b$ P.~J.~Wass,$^b$ and W.~J.~Weber$^b$ \corref{*} }

\address[cefsa]{Centro Fisica degli Stati Aggregati, 38050 Povo, Trento, Italy}
\address[utn_infn]{ Dipartimento di Fisica, Universit\`{a} di Trento, and INFN,
Gruppo di Trento, Via Sommarive 14, 38100 Povo (TN), Italy}
\cortext[*]{Corresponding author: {\it weber@science.unitn.it}}

\begin{abstract}
We present a simple analysis of the force noise associated with the
mechanical damping of the motion of a test body surrounded by a large
volume of rarefied gas. The calculation is performed considering the
momentum imparted by inelastic collisions against the sides of a cubic
test mass, and for other geometries for which the force noise could be
an experimental limitation. In addition to arriving at an accurate
estimate, by two alternative methods, we discuss the limits of the
applicability of this analysis to realistic experimental configurations
in which a test body is surrounded by residual gas inside an enclosure
that is only slightly larger than the test body itself. 

\end{abstract}

\begin{keyword}



gas damping \sep Brownian noise \sep small forces
\end{keyword}

\end{frontmatter}


\section{Introduction}
\label{intro}

Gas damping of the motion of a macroscopic test body is a potentially
sensitivity-limiting source of Brownian noise in variety of
experiments that are sensitive to very small forces. Gas damping in the
molecular flow regime is characterized by a viscous damping coefficient
$\beta_{tr} = - \frac{dF}{dV}$ in translation or $\beta_{rot} = -
\frac{dN}{d\dot{\phi}}$ in rotation, and has been found in numerous
torsion pendulum experiments to be proportional to the residual gas
pressure $p$ \cite{kenji_fused,UW_fused,superclive}. The power spectrum
of Brownian force noise associated with the molecular impacts
is related to the damping coefficient via the fluctuation-dissipation
theorem, which gives $S_F = 4 k_B T \beta_{tr}$. A quantitative study of the
gas damping and consequent force noise acting on small levitated spheres
was performed by Hinkle and Kendall \cite{hinkle_damping,
hinkle_brownian} and experimentally verified the Brownian character of
residual gas force noise. 

The order of magnitude of the translational motion gas damping
coefficient can be found easily as follows: a test body -- or test mass,
referred to here as TM -- with section $A$ that is moving with velocity
of magnitude $V$ in a gas with molecular number density $n$ (related to 
pressure and temperature by $n = \frac{p}{k_B T}$) 
will be struck, on average, by order $n A V$ more
molecules per second on the ``upwind'' side than on the ``downwind''
side. Each molecule imparts a momentum of order $m_0 v_T$ where $m_0$ is
the mass of the molecule and $v_T \equiv \sqrt{\frac{k_B T}{m_0}}$ is
the characteristic thermal velocity. This gives a translational damping
coefficient $\beta_{tr} \approx n A \times m_0 v_T \approx \frac{p A}{v_T}$. 

This general dependence is found, albeit with different prefactors, in
the analysis of gas damping force noise from damping of mirror
vibrational modes for gravitational wave interferometers and other
oscillators \cite{braginsky,saulson_prd,christian}, and for the force
noise on the cubic TM for a space gravitational wave interferometer
\cite{bonnie_LISA}. To get an exact number, we must consider not only
the momentum exchange normal to the surface, but also the role of shear
forces acting parallel to the TM surface. Additionally, we must consider
the test mass recoil from molecules leaving the surface, which is a
process that is correlated, via the conservation of particles, with the
incoming collisions.

This article presents a calculation of the gas damping coefficient, with
particular attention to the cubic TM geometry. This has been motivated
by the need for an accurate modeling of gas damping noise on the cubic
TM that serve as geodetic reference masses in the Laser Interferometry
Space Antenna (LISA \cite{big_book}) and LISA Pathfinder (LPF
\cite{nuc_phys_B}) missions, in which the TM must be free of spurious
accelerations at the level of fm/s$^2$/Hz$^{1/2}$. The dissipation
analysis for a cube is demonstrated first with a force noise evaluation
and then confirmed with direct calculation of the damping coefficient.
The force noise analysis is also extended to other simple geometries for
both translation and rotation. 

The calculation presented is straightforward and based on the simple
physics of diffuse scattering, used, for instance, in estimating
molecular flow conductances in tubes. It is of current value, however,
for several reasons. First, a quantitative calculation of gas damping is
important for the design of gravitational wave experiments, and previous
estimates appear to have underestimated this effect by roughly an order
of magnitude in noise power \cite{bonnie_LISA}. The model presented is
also easily extended to, and can be tested by, torsion pendulum small
force experiments. Finally, and most importantly, the limitations of the
applicability of this calculation -- namely the assumption of having a
TM surrounded by an infinite gas volume, which is discussed in the
conclusion -- indicate that the force noise from gas damping can be
grossly underestimated in many experiments, including LISA, which is the
subject of a more detailed experimental and numerical study \cite{us}.
The accurate calculation of damping in an infinite gas volume is thus
the necessary starting point for studying damping in tighter
geometries.

\section{Calculation}
\label{calc}

\subsection{Force noise calculation}
\subsubsection{Force noise normal and parallel to a surface element}
We can calculate the force noise spectrum per unit area -- which we define
$S_{\perp}$ and $S_{\parallel}$ for components normal and parallel to
the surface -- from the mean square fluctuations of the time average
force exerted in a time interval $T_0$. For uncorrelated
molecular impacts, which we treat as delta-functions in time, the 
time average force on a surface element $\Delta A$ can be written 
\begin{equation}
\bar{F}_{\perp} 
\equiv
\frac{1}{T_0} \int_0^{T_0} dt F_{\perp}\left( t \right) 
= \frac{1}{T_0} \sum_k m_0 \left( \vec{v}_{ik} - \vec{v}_{ok} \right)
. \hat{u}_{\perp}
\label{time_average_force}
\end{equation}
Here, $\vec{v}_{ik}$ and $\vec{v}_{ok}$ are the incoming and outgoing
velocities of the molecule involved in collision $k$ (an analogous expression
holds for the force components parallel to the surface).   

\begin{figure}[t]
\includegraphics[width=3in,keepaspectratio,clip]{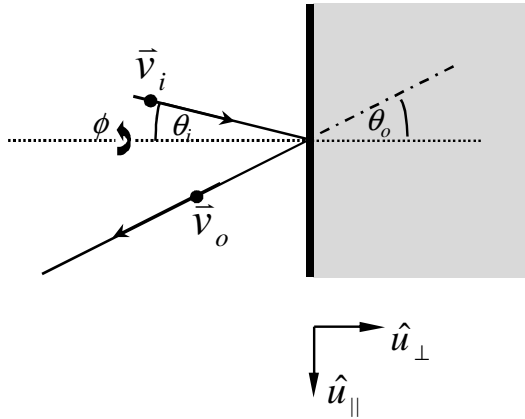}
\label{illustration}
\caption{Illustration of inelastic molecular collision with a test body 
surface, with incoming velocity $\vec{v}_i$ and subsequent reemission with
outgoing velocity $\vec{v}_o$.  Random, diffuse scattering, reemission 
with a $\cos \theta_o$ distribution is assumed in the calculation.}
\end{figure}

For white force noise, as would be expected for momentum transfer from
uncorrelated, delta-function collisions of single molecules against a
surface in the regime of molecular flow, the single-side noise spectrum
can be expressed 
\begin{equation}
S_{\perp} \Delta A = 2 T_0
\left \langle \left( \bar{F_{\perp}} - \left \langle \bar{F_{\perp}} \right \rangle
\right) ^2 \right \rangle
\label{mean_square_spectrum}
\end{equation}
with an analogous expression for $S_{\parallel}$, the force noise
per unit area parallel to a surface element.

The calculation can be simplified by considering a surface element small
enough -- and time interval short enough -- such that the probability of
more than one collision becomes negligible. In this limit -- $n v_T T_0
\Delta A \ll 1 $ -- the calculations of the expectation value of the
average force and its mean square become simple integrals over the
probability, $P \left( \vec{v}_i,\vec{v}_o \right)$, of a single
collision occuring, in the interval $T_0$, with incoming and outgoing
velocities $\vec{v}_{i}$ and $\vec{v}_{o}$,

\begin{eqnarray}
\left \langle \bar{F}_{\perp} \right \rangle  \! \! \! \!
& = &
\! \! \! \!
\frac{m_0}{T_0} \int \! \! 
d^3\vec{v}_i d^3\vec{v}_o  
P \left( \vec{v}_i,\vec{v}_o \right)
 \left( \vec{v}_{i} - \vec{v}_{o} \right)
. \hat{u}_{\perp}
\label{expectation_mean}
\\
\left \langle \bar{F}^2_{\perp} \right \rangle \! \! \! \!
& = & 
\! \! \! \!
\frac{m_0^2}{T_0^2} \int \! \!
d^3\vec{v}_i d^3\vec{v}_o 
P  \left( \vec{v}_i,\vec{v}_o \right)
 \left[ \left( \vec{v}_{i} - \vec{v}_{o} \right) 
. \hat{u}_{\perp} \right]^2
\label{expectation_mean_meansq}
\end{eqnarray}


The probability $P \left( \vec{v}_i,\vec{v}_o \right)$ can be expressed
\begin{eqnarray}
P \left( \vec{v}_i,\vec{v}_o \right) d^3\vec{v}_i d^3\vec{v}_o
=
\nonumber \\
n \left( \frac{1}{2 \pi v_T^2} \right)^{3/2}
\exp - \frac{v_{i}^2}{2 v_T^2} 
\left( \Delta A \cos \theta_{i} \, v_{i} T_0 \right)
\left( v_{i}^2 dv_{i} d\Omega_{i} \right)
\nonumber \\
\times
\frac{1}{2 \pi v_T^4}  \cos \theta_{o} \, v_{o} 
\exp - \frac{v_{o}^2}{2 v_T^2} 
\left( v_{o}^2 dv_{o} d\Omega_{o} \right)
\nonumber
\\
& &
\label{joint_prob}
\end{eqnarray}
where the velocities are expressed with magnitudes $v$ and solid angles
$\Omega$ (angles $\phi$ and $\theta$). The first term in Eqn.
\ref{joint_prob} is the probability of having a molecule, assumed to
obey a Maxwell-Boltzmann distribution with number density $n$, in the
tube of volume $dA \, \cos \theta_{i} \, v_{i} \, T_0$ that will hit the
wall element in the next time interval $T_0$ with velocity $\vec{v}_i$.
The second term corresponds to the probability, given the arrival of a
molecule, of reemission with speed $v_o$ into solid angle $\Omega_o$.
The impacts are assumed to be completely inelastic and ``memory
erasing,'' with the outgoing angle independent of the incoming angle,
and reemission is assumed to occur immediately after the impact.
Weighting by the factor $v_o \cos \theta_o$, which ensures that the
incoming and outgoing distributions are identical, is the cosine law for
diffuse scattering, which is used, with substantial experimental
verification, for estimating Knudsen diffusion in pipes and other
geometries (see, for instance, Refs. \cite{davis,mercier,
gruener_huber_prl}). We will discuss the impact of other
assumptions of the molecule - TM interaction at the conclusion of the
calculation.




We can calculate the expected average force, with $\left( \vec{v}_{i} -
\vec{v}_{o} \right) . \hat{u}_{\perp} = v_i \cos \theta_i + v_o \cos
\theta_o$ and restriction of the integral over the incoming and outgoing
velocities to the positive hemisphere ($\theta_i$, $\theta_o$ in the
range $[0,\pi/2]$):
\begin{equation}
\left \langle \bar{F}_{\perp} \right \rangle
=
m_0 \Delta A n v_T^2 = p \Delta A
\label{expect_force}
\end{equation}
with the final obvious result emerging with the ideal gas 
law, $p = n k_B T$.  

In similar fashion we can integrate to obtain the mean square average 
force, 
\begin{equation}
\left \langle \bar{F}^2_{\perp} \right \rangle
=
\frac{\Delta A}{T_0} p \left( \frac{8 m_0 k_B T}{\pi} \right)^{1/2} 
\left( 1 + \frac{\pi}{4} \right)
\label{mean_square_force}
\end{equation}

Inspection of Eqns. \ref{expect_force} and \ref{mean_square_force} 
shows that $\left \langle \bar{F}^2_{\perp} \right \rangle \gg
\left \langle \bar{F}_{\perp} \right \rangle^2$ in the studied
single-collision regime, and so Eqn. \ref{mean_square_spectrum}
simplifies to $S_{\perp} = \frac{2 T_0}{\Delta A}
\left \langle  \bar{F_{\perp}}^2 \right \rangle$, and thus
\begin{equation}
S_{\perp} = p \left( \frac{32 m_0 k_B T}{\pi} \right)^{1/2} 
\left( 1 + \frac{\pi}{4} \right) 
\label{force_normal}
\end{equation} 

We note that the ``extra'' contribution $\frac{\pi}{4}$ comes from the
cross term $\left \langle v_i \cos \theta_i v_o \cos \theta_o \right
\rangle$, which can be thought of as a correlation between the momentum
imparted normal to the surface, which is always positive, by the arrival
and subsequent reemission of a molecule. The remaining term -- unity in
parentheses in Eqn. \ref{force_normal} -- comes from equal contributions
of the incoming and outgoing molecules. 

If, in alternative, we were to consider the incoming and outgoing
particle fluxes as uncorrelated processes, which would be true for
molecules that stick to the surface for a long duration -- much longer
than the reciprocal of the frequency where we consider the force noise
-- before reemission, the contribution $\frac{\pi}{4}$ disappears.
Another interesting case is that of elastic collisions with specular
reflection, which is equivalent to replacing the outgoing distribution
with a delta function for the specular reflection of the incoming
velocity, the factor $\left ( 1 + \frac{\pi}{4} \right )$ is replaced
by a factor 2.

We can perform a similar calculation for either of the force 
components parallel to the surface.  For this we replace, in
evaluating Eqn. \ref{expectation_mean_meansq}, the 
momentum exchange normal to the surface with one of the components
parallel to the surface \footnote[1]{The resulting noise is the 
 same for the other parallel component, with $\sin \phi$
replacing $\cos \phi$ in the calculation}, 
\begin{equation}
m_0 \left( \vec{v}_{i} - \vec{v}_{o} \right)
. \hat{u}_{\parallel} 
= v_i \sin \theta_i \cos \phi_i - v_o \sin \theta_o \cos \phi_o
\label{component_parallel}
\nonumber
\end{equation}
For this parallel component, the expectation average force is zero.
Also, in the mean square of the average force, there is no effect 
of correlation between the incoming and outgoing molecules, as the
azimuth angles are independent of one another and both
have zero mean momentum exchange parallel to the surface.  We thus 
obtain 
\begin{equation}
S_{\parallel} = p \left( \frac{8 m_0 k_B T}{\pi} \right)^{1/2}
\label{force_parallel}
\end{equation} 

Considering the alternative cases discussed above, there is no
difference in the case of sticking molecules, as the incoming 
and outgoing processes are uncorrelated.  In the case of specular
reflection, the velocity parallel to the surface element is conserved
and thus $S_{\parallel} = 0$.   

Finally, we note that in the diffuse scattering hypothesis there is no
cross-correlation between the different components of the force noise,
given the azimuthally random reemission of molecules.

\subsubsection{Total force and torque noise on a cubic 
TM and other geometries}
The force noise normal (Eqn. \ref{force_normal}) and parallel (Eqn.
\ref{force_parallel}) to a unit surface element can be integrated over
the surface of a test body to give the total force along a chosen axis.
For a cubic test mass, this calculation is particularly simple, with the
relevant force noise component per area ($S_z$ for the $z$ axis) uniform
on each face, given by $S_{\perp}$ on the two faces ($Z$) normal to $z$
and by $S_{\parallel}$ on the four faces ($X$ and $Y$) parallel to $z$.
For a cube of side length $s$, we obtain a total force noise 

\begin{eqnarray}
S_F 
& = &  
2 \int_{Z} dx dy S_{\perp}
+
2 \int_{X} dy dz S_{\parallel}
+
2 \int_{Y} dx dz S_{\parallel}
\nonumber 
\\ 
& = & p s^2 \left( \frac{512 \, m_0 k_B T}{\pi} \right) ^{1/2} 
\left( 1 + \frac{\pi}{8} \right)
\label{force_noise}
\end{eqnarray}

We can also evaluate the torque noise along a chosen axis, which is of
relevance to torsion pendulum experiments. Considering the torque along
the $z$ axis, $N_z = x F_y - y F_x$, we integrate the torque noise per
area for every surface element, $S_{N_z} = x^2 S_y + y^2 S_x$, over the
faces of the cube. For the $Z$ faces, only shear forces, in $x$ and $y$
contribute to the torque. The four lateral faces each have contributions
from both shear, with armlength $\frac{s}{2}$, and normal forces, with
an armlength that varies across the cubic face. Integrating, we obtain
\begin{eqnarray}
S_N & = & 2 \int_{X} dy dz \left[  y^2 S_{\perp} 
 + \left( \frac{s}{2} \right)^2 S_{\parallel} \right]
\nonumber
\\
& & + 2 \int_{Y} dx dz \left[  x^2 S_{\perp} 
 + \left( \frac{s}{2} \right)^2 S_{\parallel} \right]
\nonumber
\\
& & + 2 \int_{Z} dx dy 
\left[ x^2 S_{\parallel} +  y^2 S_{\parallel} \right]
\nonumber
\\
& = & \frac{s^4}{3} S_{\perp} +  \frac{4 s^4}{3} S_{\parallel} 
\nonumber
\\
& = & p s^4 
\left( \frac{32 \, m_0 k_B T}{\pi} \right)^{1/2}  
\left( 1 + \frac{\pi}{12} \right)
\label{torque_noise}
\end{eqnarray}

The calculation of the total force and torque noise has also been
performed for a sphere and a right cylinder, by straightforward
integration of Eqns. \ref{force_normal} and \ref{force_parallel} over
the TM surface (results in Table 1). We note that the rotational damping
(or, equivalently, the torque noise) on a sphere or cylinder, the latter
along its axis of symmetry, is due entirely to the shear forces and thus
would be zero in a model with elastic molecular collisions. 

\begin{table*}[t]
\label{shapes}
\centering
\begin{tabular}{ m{1in}c m{1in}c  m{1in}c     }

\hline \\ 
shape  

&  $\beta_{tr} / \left[ p / v_T \right]$ 

& $\beta_{rot} / \left[ p / v_T \right]$ \\	\\

\hline \\ [1ex]
\includegraphics[width=0.75in,keepaspectratio,clip]{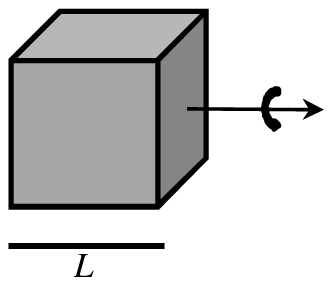}
&  
$s^2 \left( \frac{32}{\pi} \right)^{1/2} \left( 1 + \frac{\pi}{8} \right) \:\:$

&
$s^4 \left( \frac{2}{\pi} \right)^{1/2} \left( 1 + \frac{\pi}{12} \right)$  \\ \\
\includegraphics[width=0.75in,keepaspectratio,clip]{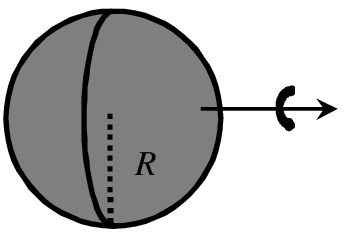}
 & 
$\pi R^2 \left( \frac{128}{9 \pi} \right)^{1/2} \left( 1 + \frac{\pi}{8} \right) \:\:$  
&
$\pi R^4 \left( \frac{32}{9 \pi} \right)^{1/2} $ \\ \\

\includegraphics[width=0.75in,keepaspectratio,clip]{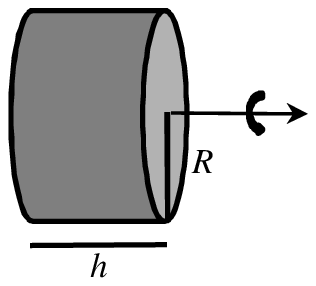}
 & 
$\pi R^2 \left( \frac{8}{\pi} \right)^{1/2} 
\left( 1 + \frac{h}{2R} + \frac{\pi}{4} \right) \:\:$   
&
$\pi R^4 \left( \frac{1}{2 \pi} \right)^{1/2} 
\left( 1 + \frac{2h}{R}  \right)$  \\ \\

\includegraphics[width=0.75in,keepaspectratio,clip]{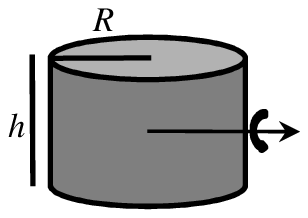}
 & 
$\pi R^2 \left( \frac{2}{\pi} \right)^{1/2} 
\left[ 1 + \frac{3h}{2R} \left(1  + \frac{\pi}{6} \right) \right] \:\:$   
&
$\pi R^4 \left( \frac{1}{2 \pi} \right)^{1/2} \times $ \\ 

& & 
$  
\left[ 1 + \frac{\pi}{4} + \frac{h}{R}  + \frac{h^2}{2R^2} 
+ \frac{h^3}{4 R^3} \left( 1 + \frac{\pi}{6} \right) \right] 
$  
\\ [1ex]
\hline
\end{tabular}
\caption{Calculated translational and rotational damping coefficients
for different test bodies with cubic, spherical, and cylindrical shapes,
along the axes indicated in the figures at left. Note that the damping
coefficients are related to the force and torque noise spectra via the
fluctuation-dissipation formulas, $S_F = 4 k_B T \beta_{tr}$ and $S_N =
4 k_B T \beta_{rot}$.}
\end{table*}

We note that the results of Table 1 have been confirmed to roughly the 
percent level, for these three geometries and in both translation and 
rotation, by a numerical simulation that traces the impulses of a single 
molecule as it impacts the test mass inside a much larger volume, with
the reemission distribution governed by the diffuse scattering statistics
of Eqn. \ref{joint_prob} \cite{hinkle_foot}.  The simulation technique 
is described in Ref. \cite{us}.   

\subsection{Direct calculation of viscous damping coefficient for a cube}
As a simple crosscheck for the force noise calculated for a cube in the
previous section, we can directly calculate the viscous damping
coefficient $\beta$ by calculating the force on a cubic test body moving
with velocity $\vec{V}=V_{\perp} \hat{u}_{\perp}$ with respect to the
surrounding gas. Damping will occur both due to the molecules which hit
the test mass on all faces -- with a velocity distribution that becomes
asymmetric, in the test mass reference frame, due to $V_{\perp}$ -- and
due to the reemitted molecules -- which, though emitted with the same
velocity distribution on all sides of the test mass, still produce a net
force, as more molecules will be emitted from the upwind side.

The expectation average force on a surface element on the upwind or
downwind faces -- the faces perpendicular to the test mass motion --
can be calculated, as in Eqn. \ref{expectation_mean}, by integrating
the momentum exchange $m_0 \left( v_{i \perp} - v_{o \perp} \right)$ 
over the joint probability function, $P_{V_{\perp}}$, which assumes 
an asymmetry in the reference frame of the moving test mass, 
\begin{eqnarray}
P_{V_{\perp}} \left( \vec{v}_i,\vec{v}_o \right) d^3\vec{v}_i d^3\vec{v}_o
=
\nonumber \\
n \left( \frac{1}{2 \pi v_T^2} \right)^{3/2}
\exp - \frac{ \left( v_{i{\perp}}+V_{\perp} \right)^2 + v_{i \parallel}^2}{2 v_T^2} 
\nonumber \\
\times 
\left( \Delta A v_{i\perp} T_0 \right)
\left( v_{i}^2 dv_{i} d\Omega_{i} \right)
\nonumber \\
\times
\frac{1}{2 \pi v_T^4}  v_{o \perp} 
\exp - \frac{v_{o}^2}{2 v_T^2} 
\left( v_{o}^2 dv_{o} d\Omega_{o} \right)
\nonumber
\\
& &
\label{joint_prob_V}
\end{eqnarray}

The integrals can be simplified in the limit of $V_{\perp} \ll v_T$ and
neglecting terms of order $V_{\perp}^2$, to yield, respectively, 
for the downwind and upwind faces, 
\begin{equation}
\left \langle \bar{F}_{\perp} \right \rangle 
= 
p \Delta A 
\left[ \pm 1  - V_{\perp} \left( \frac{2 m_0}{\pi k_B T} \right)^{1/2} 
\left( 1  + \frac{\pi}{4} \right) 
\right]
\label{expect_force_V_perp}
\end{equation}
Again, had we assumed elastic collisions, which essentially multiplies
the force of the incoming molecules by two, the factor 
$\left( 1  + \frac{\pi}{4} \right)$ would become 2, which would produce 
the same result obtained by Christian \cite{christian} for the difference
in pressure on two sides of a membrane.  

We can similarly calculate the expected force component 
for a surface element on the faces parallel to the TM
motion, following Eqn. \ref{component_parallel}, which yields 
\begin{equation}
\left \langle \bar{F}_{\parallel} \right \rangle 
= -
p \Delta A 
V_{\parallel}
\left( \frac{m_0}{2 \pi k_B T} \right)^{1/2}  
\label{expect_force_V_par}
\end{equation}
 
Equations \ref{expect_force_V_perp} and \ref{expect_force_V_par} can be
integrated -- actually a simple multiplication by the surface area $s^2$
-- over the six faces of the cube. The leading term in Eqn.
\ref{expect_force_V_perp} cancels out upon summing the upwind and
downwind faces, to yield a pure velocity damping force, 
\begin{equation}
\beta 
\equiv
 - \frac{d F_x}{d V_x} 
= p s^2 \left( \frac{32 m_0} {\pi k_B T} \right)^{1/2} 
\left( 1 + \frac{\pi}{8} \right)  
\label{beta_trans}
\end{equation}

This agrees with the value of $\beta$, given in Table 1 and calculated
from the force fluctuations and thus, as expected, agrees with the
fluctuation-dissipation theorem.

\section{Discussion}
\label{discussion}

The expressions derived here for the gas damping on a cubic test mass
are thus confirmed by two independent calculations. Both calculations of
course rest on the assumptions of diffuse scattering of molecules along
the test mass surface. This would be fairly easy to test, for the cube
(or similar plate) and cylinder geometries for the pressure dependence
of the rotational damping coefficient in a torsion pendulum. We also
note that measuring the rotational damping for a cylinder is a direct
test of the diffuse scattering hypothesis, as elastic scattering would
give no rotational damping. 

Applying Eqn. \ref{force_noise} to a 46 mm cubic TM, as currently
employed for LISA, the expected force noise from molecular impacts 
in an infinite gas volume would be 

\begin{eqnarray}
S_F^{1/2}  
= 0.75 \, \mathrm{fN / Hz^{1/2}} \, \times \, 
\nonumber \\
\left( \frac{p}{1 \: \mathrm{\mu Pa}} \right)^{1/2}
\left( \frac{s}{46 \: \mathrm{mm}} \right)^{2}
\left( \frac{m_0}{30 \: m_p} \right)^{1/4}
\label{LISA}
\end{eqnarray}

Two points are important here. The first is the magnitude of this term,
which, for a given pressure and test mass dimension, is roughly three
times larger than that obtained in a previous analysis
\cite{bonnie_LISA}. Given the slow scaling with pressure, $S_F^{1/2}
\sim p^{1/2}$, more ambitious gravitational wave missions, such as BBO
\cite{bbo} and DECIGO \cite{decigo}, which require lower acceleration
noise, will have to reach much lower pressures than the $10^{-6}$ Pa
value quoted here, or employ much larger test masses, with a larger mass
more than offsetting the larger surface area in Eqn. \ref{LISA}. In
general, for given TM density, the resulting acceleration noise is given
by $S_a^{1/2} \sim p^{1/2} s^{-2} \sim p^{1/2} M^{-2/3}$ for a TM of
mass $M$ and dimension $s$. 

The second consideration concerns the applicability of this calculation
to frequently encountered experimental conditions in which the test mass
is surrounded by an enclosure which is only slightly larger than the
test mass itself, or more generally is very close to the surrounding
apparatus. This is certainly true for LISA, where the 46 mm cubic test
mass is enclosed, with a gap of several mm, by the walls of a
surrounding electrostatic position sensor \cite{spie_sensor}. This is
also true for terrestrial gravitational wave observatories such as
Advanced LIGO \cite{adv_ligo}, where suspended TM are in close proximity
to similar suspended masses used for actuation and thermal compensation.
Similar physics of gas damping in a restricted geometry is fundamental
to recent studies of ``squeeze damping'' in MEMS
resonators\cite{suijlen}.

For such cases, the
assumption, used in the calculations presented here, of being able to
calculate the force noise from the probability of a single impact --
which is reasonable only if that impact is uncorrelated with previous
and future impacts of the same molecule -- collapses. In a tightly
restricted geometry, a molecule emitted from a TM surface will,
with high probability, strike the opposing enclosure wall only to then
return to strike the same TM surface, imparting momentum with the
same sign as the previous impact. Random walking, by molecular
diffusion, from one side of the test mass to the other, will require many
collisions and create a grouping of correlated collisions with
force impulses with the same sign. This increases force noise, as the
momentum imparted does not average out as quickly as for
uncorrelated collisions in an infinite gas volume. 

The random walk process down the narrow channels between test mass and
enclosure is the same that determines the flow impedance in a tube.
Given that test mass motion, inside a tight enclosure, will require gas
flow around the test mass, this impedance creates a pressure head, thus
giving an alternate picture of the velocity-dependent force, or 
dissipation, that must accompany the increased force noise. These
constrained volume effects are studied both experimentally and with
numerical simulations, in a recently published parallel study \cite{us},
which, in addition to obtaining the force and torque noise formulas
derived here in the limit of very large volume, demonstrates
significantly increased damping in the case of gaps that are smaller
than the TM itself.




\begin{thebibliography}{00}
\bibitem{hinkle_foot}
We note that our analytic results, and their confirmation by numerical
simulation, present a slight discrepancy compared to the result of
\cite{hinkle_brownian} (Eqn. 6 in that article) for the translational
gas damping of a sphere, in the diffuse scattering hypothesis and
apparently including the correlation between incoming and outgoing
molecules. That result has a term $\frac{\pi}{16}$ in place of the
$\frac{\pi}{8}$ given in Table 1 here.
\end{thebibliography}
\end{document}